\documentclass[a4paper]{jpconf}
\usepackage{graphicx}
\begin{document}
\title{LHC/ILC Interplay in SUSY Searches}

\author{G.\ Moortgat-Pick}

\address{
IPPP, Institute for Particle Physics Phenomenology, 
University of Durham, Durham DH1 3LE, UK
}

\ead{g.a.moortgat-pick@durham.ac.uk}

\begin{abstract}
Combined analyses at the Large Hadron Collider 
and at the International Linear Collider are important to reveal 
precisely the new physics model as, for instance, supersymmetry. 
Examples are presented where ILC results as input for LHC analyses
could be crucial for the identification of signals as well as 
of the underlying model.
The synergy of both colliders leads also to rather accurate 
SUSY parameter determination and
powerful mass constraints even
if the scalar particles have masses in the multi-TeV range.
\end{abstract}

\vspace{-.5cm}
\section{LHC/ILC interplay in the gaugino/higgsino sector}
Supersymmetry (SUSY) is one of the best-motivated candidates for physics
beyond the Standard Model (SM).
If experiments at future accelerators, the Large Hadron
Collider (LHC) and the International Linear Collider (ILC), discover SUSY 
they will also have to determine precisely the 
underlying  SUSY-breaking scenario.
Methods  to derive the SUSY parameters
at collider experiments have been worked out, for instance
in~\cite{Tsukamoto:1993gt,Choi:1998ut,Choi:2001ww,Desch:2003vw,Bechtle:2004pc}.

\vspace{-.3cm}
\subsection{Mass predictions for the heavy gauginos/higgsinos}
In ~\cite{Desch:2003vw,Weiglein:2004hn} it has been studied in detail for a
representative SUSY scenario SPS1a~\cite{sps} how results at the ILC
can benefit from LHC input and vice versa.

\vspace{.1cm}
\noindent{\it LHC analysis:} In most cases the masses  
of the Susy particles can only be studied by analysing complicated 
decay chains, like
\begin{equation}
\tilde{q}_L\to \tilde{\chi}^0_2 q \to \tilde{\ell}^{\mp}_R \ell^{\pm} q\to 
\tilde{\chi}^0_1 \ell^{\mp} \ell^{\pm} q,
\label{eq_chain}
\end{equation}
which might be difficult to resolve.  
The precise reconstruction of the states in
the decay chains requires in particular the knowledge of the mass of the
lightest Susy particle (LSP), which is often assumed to be stable.
In SPS1a the second lightest neutralino can be identified in the
opposite sign-same flavour signal (OS-SF) with an uncertainty of about
$\delta m_{\tilde{\chi}^0_2}=4.7$~GeV. To measure the heavier 
gauginos/higgsinos is extremely challenging due to mass degeneration.

\vspace{.1cm}
\noindent{\it ILC analysis:}
Precise simulations for the mass measurements of the sleptons and the
light charginos and neutralinos show that an accuracy of much less than 1~GeV
can be achieved  at the ILC with $\sqrt{s}=500$~GeV.  
Particularly interesting is the
high accuracy in the determination of $m_{\tilde{\chi}^0_1}$ with
$\delta(m_{\tilde{\chi}^0_1})=0.05$~GeV from $\tilde{e}_R$ decays, but
also the accuracy $\delta (m_{\tilde{\chi}^{\pm}_1})=0.55$~GeV and
$\delta (m_{\tilde{\chi}^{0}_2})=1.2$~GeV are important. 

It has been assumed that the cross sections for $\tilde{\chi}^+_1\tilde{\chi}^-_1$ and 
$\tilde{\chi}^0_1\tilde{\chi}^0_2$ have been precisely measured at different 
energies $\sqrt{s}=400$, $500$~GeV and for different beam polarizations 
$(P_{e^-},P_{e^+})=(\pm 80\%, \mp 60\%)$~\cite{Moortgat-Pick:2005cw}.

\begin{figure}[h]
\includegraphics[width=14pc]{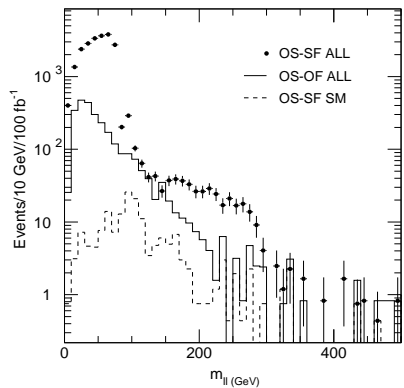}\hspace{2pc}%
\begin{minipage}[b]{20pc}
\caption{\label{fig-mchi04}  
 The invariant mass spectrum of 
the heavy neutralino/chargino decay chains.
The dilepton opposite-sign--same-flavour lepton edge of 
$\tilde{\chi}^0_4$ is the edge between
200~GeV$<m_{ll}<$400~GeV~\cite{Desch:2003vw,Weiglein:2004hn}.
\vspace{1cm}
}    
\end{minipage}
\end{figure}

\vspace{.2cm}
\noindent{\it LHC/ILC interplay analysis:}
Putting together all results from LHC and ILC
allows to determine precisely the fundamental
electroweak parameters in the 
minimal supersymmetric standard model (MSSM), 
$M_1$, $M_2$, $\mu$ and $\tan\beta$, without 
any assumption on the SUSY breaking scheme.  The masses of the
heavy neutralinos and charginos can now be predicted.
Concerning again studies at the
LHC such mass predictions from the ILC analysis lead to an increase of
statistical sensitivity, which could be crucial for the search for
statistically marginal signals:  
Together with a precise knowledge on
the mass of the lightest stable particle $\tilde{\chi}^0_1$ and the
light slepton masses measured at the ILC, the mass predictions lead to
a clear identification of the dilepton edge from the
$\tilde{\chi}^0_4$ decay chain, followed by precise mass measurements
of these heavy particles, cf. Fig.~\ref{fig-mchi04}.
Measuring the heavy particles right at the predicted masses provides
an important check of the underlying Susy model. 

\subsection{Model distinction between MSSM and NMSSM}
Such an important model check has been studied in detail
in~\cite{MoortgatPick:2005vs}, where two basic supersymmetric models,
the MSSM and the
next-to-minimal supersymmetric standard model (NMSSM) have been
analyzed.  The NMSSM \cite{nmssm} is the simplest extension of the
MSSM by an additional Higgs singlet field. The corresponding
additional fifth neutralino may significantly change the phenomenology
of the neutralino sector due to the suppressed coupling to ordinary
gauge bosons of the singlino admixture~\cite{singlino}.

In this case study a scenario has been presented where
all kinematically accessible neutralinos and charginos have
similar masses and almost identical cross sections,
within experimental errors, in MSSM and NMSSM. 
Although the second lightest neutralino in the NMSSM has a significant
singlino component, the models cannot be distinguished
by the experimental results at the LHC or at
the ILC$_{500}$ with $\sqrt{s}=500$ GeV alone 
if only measurements of masses, cross sections and
gaugino branching ratios are considered. Also 
the Higgs sector does not allow the identification of the NMSSM.
Precision measurements of the neutralino branching ratio 
into the lightest Higgs particle and of  
the mass difference between the lightest and next-to-lightest 
SUSY particle
may give first evidence for the SUSY model but are  
difficult to realize in our case. 
Therefore the identification of the
underlying model requires precision measurements of the
heavier neutralinos by combined analyses of LHC and ILC.

Although the neutralinos $\tilde{\chi}^0_{2,3}$ have 
significant singlet components of about $>42\%$ in the
NMSSM, 
the masses of the accessible light neutralinos and charginos,
as well as the production cross sections, lead to identical values in
the two models within experimental errors.  

\begin{figure}[h]
\begin{minipage}{17.5pc}
\includegraphics[width=17.5pc,height=13.pc]{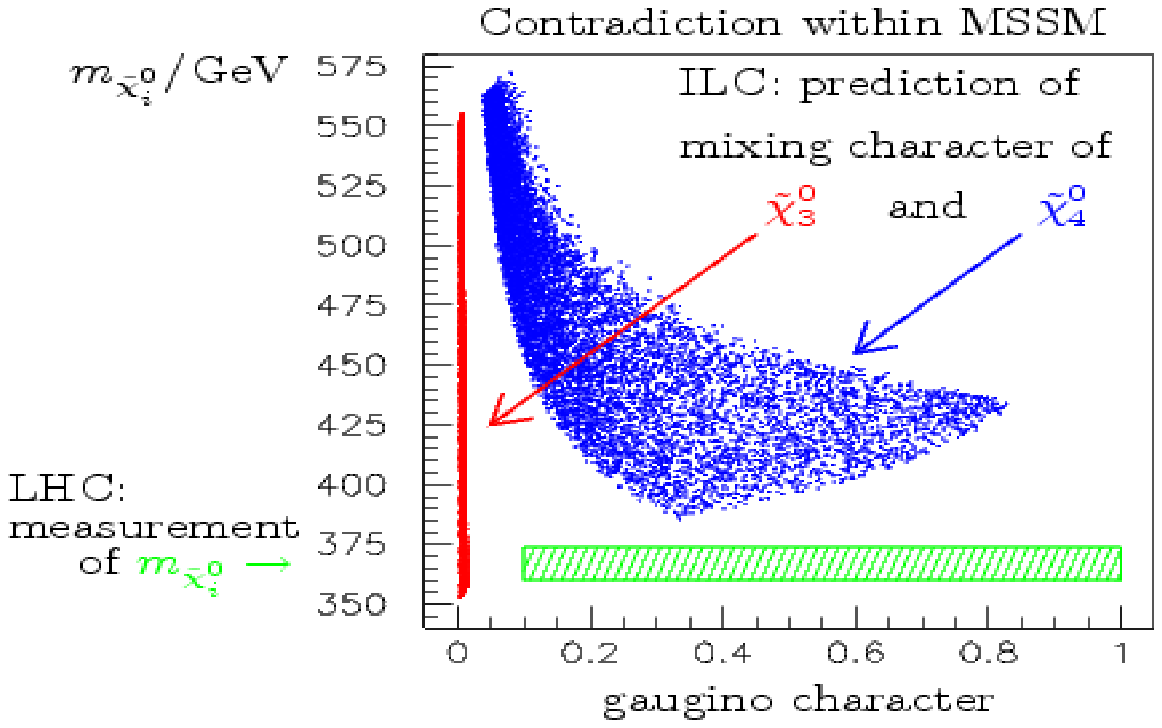}
\vspace*{-.8cm}
\caption{\label{fig-nmssm}
Predicted masses and gaugino admixture for the heavier neutralinos
$\tilde{\chi}^0_3$  and $\tilde{\chi}^0_4$ within the 
parameter ranges consistent at the ILC$_{500}$ analysis
in the MSSM and a measured mass $m_{\tilde{\chi}^0_i}=367\pm 7$~GeV
of a neutralino with sufficiently high gaugino admixture 
in cascade
decays at the LHC. We took a lower bound of a detectable gaugino admixture
of about 10\%~\cite{MoortgatPick:2005vs}.}
\end{minipage}\hspace{2.5pc}%
\begin{minipage}{17.5pc}
\includegraphics[width=17.5pc,height=12pc]{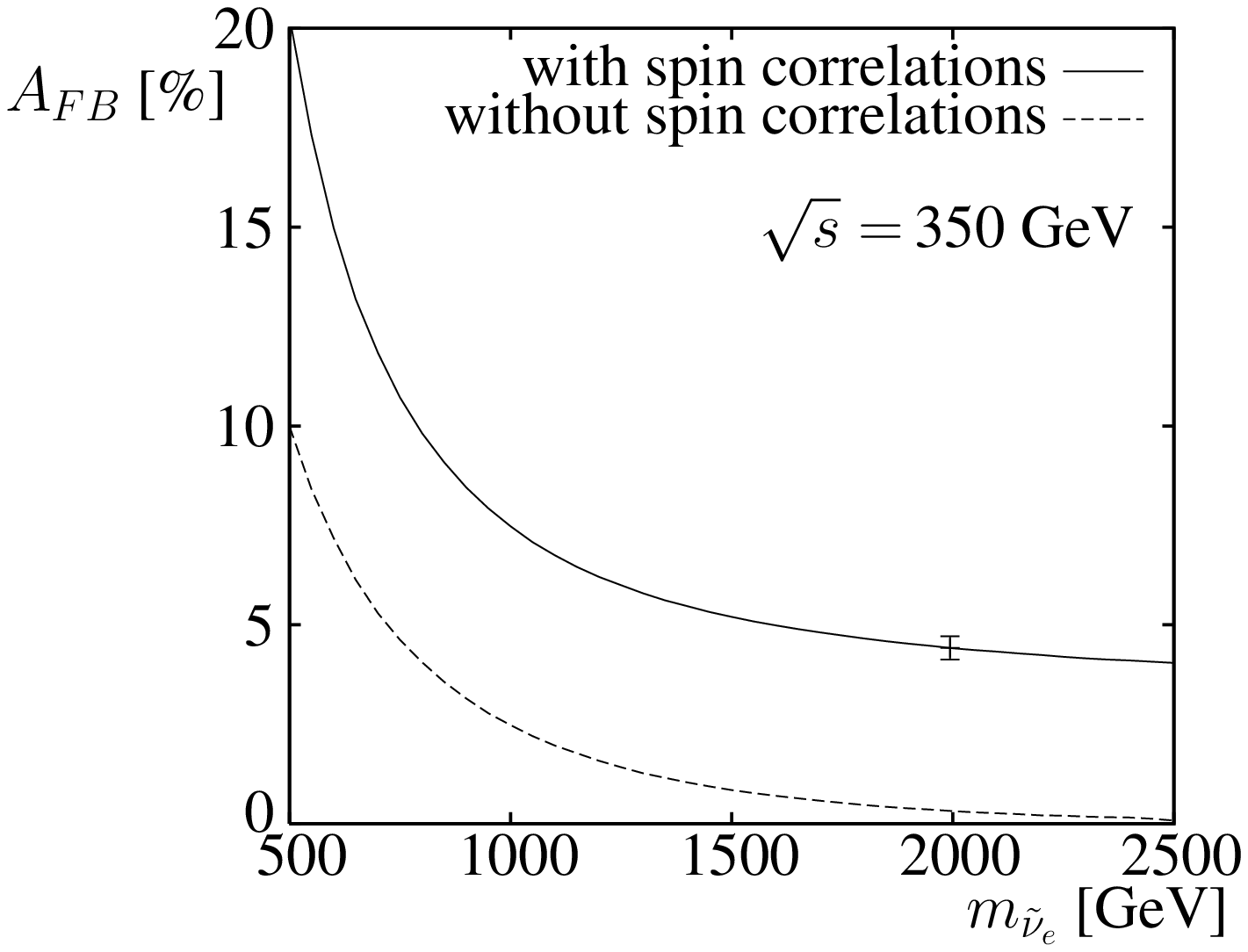}
\caption{\label{fig-msneu}
Forward--backward asymmetry of $e^-$ in
the process $e^+e^- \to \tilde{\chi}^+_1 \tilde{\chi}^-_1$,
$\tilde{\chi}^-_1 \to \tilde{\chi}^0_1 \ell^- \bar{\nu}$, 
shown as a function of
$m_{\tilde{\nu}_e}$.  
For nominal value of $m_{\tilde{\nu}_e}=1994$ GeV the
expected experimental errors are shown.  For
  illustration only, the dashed 
line shows that neglecting spin
correlations would lead to a completely 
wrong interpretation of the experimental data~\cite{Desch:2006xp}.
}
\end{minipage} 
\end{figure}

\vspace{.2cm}
\noindent{\it LHC/ILC analysis:} As described in the previous section,
the interplay analysis at both colliders allows to determine the
fundamental parameters $M_1$, $M_2$, $\mu$ and $\tan\beta$ with high precision.
The parameters lead within the assumed experimental 
uncertainty to predictions for the heavy 
neutralinos and their mixing character:
the MSSM predicts an almost pure
higgsino-like state for $\tilde{\chi}^0_3$
and a mixed gaugino-higgsino-like $\tilde{\chi}^0_4$, see Fig.~\ref{fig-nmssm}.
However, such a prediction of the mixing character contradicts
the outcome of the LHC, where only neutralinos with a sufficiently 
high gaugino admixture can be resolved: $m_{\tilde{\chi}^0_3}=367\pm 7 $~GeV.
Such a contradiction leads obviously to the correct identification of the
supersymmetric model.

\section{LHC/ILC results for unravelling multi-TeV scalar fermions}
Scenarios where the squark and slepton masses
are very heavy (multi-TeV range) are particularly challenging. 
In most studies to determine the fundamental 
SUSY parameters, it has been assumed that the
masses of the virtual scalar particles are already known. 
In the case of heavy scalars such assumptions, however, cannot be applied. 
It has been shown in~\cite{Moortgat-Pick:1999ck,Desch:2006xp} that via 
exploiting spin effects~\cite{Moortgat-Pick:1998sk},
useful indirect bounds for the mass of the heavy virtual particles could be 
derived from forward--backward asymmetries of the final lepton 
$A_{\rm FB}(\ell)$. In~\cite{Desch:2006xp} a case study
with $\sim$ 2 TeV scalar particles sector is discussed:
$m_{\tilde{\chi}^{\pm}_{1,2}}=117, 552$~GeV, 
$m_{\tilde{\chi}^{0}_{1,2,3,4}}=59, 117, 545, 550$~GeV,
$m_{h}=119$~GeV, $m_{\tilde{g}}=416$~GeV and 
$m_{\tilde{\nu}_e}$, $m_{\tilde{e}_{\rm {R,L}}}\sim 2$~TeV  
$m_{\tilde{q}_{\rm {R,L}}}\sim 2$~TeV.

\vspace{.1cm}
\noindent {\it Analysis at the LHC:} All
squarks in this scenario are kinematically accessible at the LHC. 
However, since $m_{\tilde{q}_{\rm L,R}}\sim 2$~TeV, precise mass
reconstruction will be difficult.
 Since the gluino is rather light in this scenario, 
several gluino decay channels can be exploited. The
largest branching ratio is 
a three-body  decay  into
neutralinos, $BR(\tilde{g}\to \tilde{\chi}^0_2 b
\bar{b})\sim 14\%$, 
followed by a subsequent three-body leptonic neutralino decay
$BR(\tilde{\chi}^0_2\to 
\tilde{\chi}^0_1 \ell^+ \ell^-)$, $\ell=e,\mu$ of about 6\%. 
The mass difference between the
two light neutralino masses can be measured from the dilepton edge
with an uncertainty of about
$\delta(m_{\tilde{\chi}^0_2}-m_{\tilde{\chi}^0_1})
\sim 0.5~\mathrm{ GeV}$~\cite{Weiglein:2004hn,Kawagoe:2004rz}.
It is expected to reconstruct the gluino mass 
with  a relative uncertainty of $\sim$2\%~\cite{Gjelsten:2005aw}.

\vspace{.1cm}
\noindent {\it Analysis at the ILC:} 
At the first stage of the ILC, 
$\sqrt{s}\le 500$~GeV, only light charginos and
neutralinos are kinematically accessible. However, 
in this scenario
the neutralino sector is characterized by very low production cross
sections, below 1~fb, so that it 
might not be fully exploitable~\cite{Desch:2006xp}.
Only the chargino pair production process has high rates
and $\sqrt{s}=350$ and $500$~GeV are used.
The chargino mass can be measured in
the continuum, with an error of about 
$ 0.5$~GeV, optimized via threshold scans to
$m_{\tilde{\chi}^{\pm}_1}=117.1 \pm 0.1 ~\mathrm{ GeV}$~\cite{Martyn:1999tc}.

The mass of the lightest neutralino
$m_{\tilde{\chi}^0_1}$ can be derived,
either from the lepton energy distribution ($BR(\tilde{\chi}^{-}_1\to
\tilde{\chi}^0_1 \ell^-
\bar{\nu}_\ell)\sim 11\%$
or from the invariant mass distribution of the two jets
 ($BR(\tilde{\chi}^-_1\to \tilde{\chi}^0_1 q_d \bar{q}_u)\sim 33\%$: 
$m_{\tilde{\chi}^{0}_1}=59.2 \pm 0.2~\mathrm{ GeV}$~\cite{Martyn:1999tc}.
Together with the information from the LHC
a mass uncertainty can be assumed for the second lightest neutralino of about
$m_{\tilde{\chi}^{0}_2}=117.1 \pm 0.5 ~\mathrm{ GeV.}$
The dominant SM background is $W^+W^-$ production. For the
semileptonic (slc) final state, this background can be efficiently
reduced from the reconstruction of the hadronic invariant mass. An overall 
selection efficiency of 50\% in the fully leptonic and 
semileptonic final states has been estimated. Statistical uncertainties for the
cross section and $A_{\rm FB}$(see ~\cite{Desch:2006xp}) based on 
${\cal L}=200$~fb$^{-1}$ in each polarization
configuration, $(P_{e^-},P_{e^+})=(-90\%,+60\%)$ and
$(+90\%,-60\%)$, and a relative uncertainty in the polarization of
$\Delta P_{e^\pm}/P_{e^\pm}=0.5\%$~\cite{Moortgat-Pick:2005cw} have been 
taken into account.

\vspace{.2cm}
\noindent {\it Parameter determination in LHC/ILC interplay:}\\
The underlying SUSY parameters are determined 
in two steps:\\[-1.2em] 
\begin{itemize}
\item[a)]
only the masses of
$\tilde{\chi}^{\pm}_1$, $\tilde{\chi}^0_1$, $\tilde{\chi}^0_2$ and the
chargino pair production cross section, including the full leptonic
and the semileptonic decays have been used as observables. 
A four-parameter fit has been 
applied for the parameters $M_1$, $M_2$,
$\mu$ and $m_{\tilde{\nu}_e}$ for fixed values of $\tan\beta=5$, 10, 15, 20,
25, 30, 50 and 100. Due to the strong correlations among 
parameters~\cite{Desch:2006xp},
fixing $\tan\beta$ is necessary. 
A $\chi^2$ test has been performed and one obtains:
\begin{eqnarray*}
&&59.4  \le M_1 \le 62.2~\mathrm{ GeV},\quad
 118.7  \le M_2 \le 127.5~\mathrm{ GeV},\quad  
 450  \le \mu \le 750~\mathrm{ GeV},
\\&&
1800 \le m_{\tilde{\nu}_e}\le 2210~\mathrm{ GeV}.
\end{eqnarray*}
\item[b)] 
the leptonic forward--backward asymmetry has been included as additional 
observable, exploiting 
full spin correlations between production and
 decay~\cite{Moortgat-Pick:1998sk}.  
Only the semileptonic and purely
 leptonic decays were considered.  The $SU(2)$ relation between the
 two virtual masses $m_{\tilde{\nu}}$ and $m_{\tilde{e}_{\rm L}}$ has
 been applied.

The multiparameter 
fit strongly improves the results. No assumption on
$\tan\beta$ has to be made. 
The results are
\begin{eqnarray*}
&& 59.7\le M_1\le 60.35~\mathrm{ GeV},\quad
119.9\le M_2 \le 122.0~\mathrm{ GeV},\quad
500\le \mu \le 610~\mathrm{ GeV},\\&& 
14\le \tan\beta  \le 31,\quad
1900\le m_{\tilde{\nu}_e} \le 2100~\mathrm{ GeV}.
\end{eqnarray*}

Mainly the constraints for the mass $m_{\tilde{\nu}_e}$
are improved by a factor of about $2$, see Fig.~\ref{fig-msneu}, and 
for gaugino mass
parameters $M_1$ and $M_2$ by a factor of about $5$.
The higher masses are predicted to be within the ranges
$506 < m_{\tilde{\chi}^0_3} < 615 \mbox{ GeV}$,  
$512 < m_{\tilde{\chi}^0_4} < 619 \mbox{ GeV}$,  
$514 < m_{\tilde{\chi}^{\pm}_2} < 621\mbox{ GeV}.$
\end{itemize}
Scenarios with heavy scalar particles 
are challenging  for determining the MSSM parameters. 
The forward--backward asymmetry is a powerful observable and
strongly dependent on
the mass of the exchanged heavy particle. If the $SU(2)$ constraint
is applied, the slepton masses can be determined to a precision of
about 5\% for masses around 2~TeV at the ILC running at 500~GeV.  
One should note that the analysis is performed entirely at the EW scale
and without any reference to the underlying SUSY-breaking mechanism.

\vspace{-.2cm}
\section{Conclusions}
\label{chap4}
LHC/ILC synergy can be crucial for resolving SUSY signals of heavy
particles, determining the underlying SUSY model and fixing the
fundamental parameters. Even challenging scenarios where neither the
LHC nor the ILC alone can provide the needed data, the combined
analysis of both colliders is expected to be successful.  The high
precision measurements allow to determine the parameters very
accurately, leading to
powerful mass predictions of the heavier particles. Even scenarios with 
multi-TeV sparticles can be resolved. Such predictions
might become crucial for outlining the higher energy steps of the ILC.

\vspace{-.2cm}
{\small
\ack
GMP would like to thank K. Desch, H. Fraas, F. Franke, S. Hesselbach, 
J. Kalinowski, M. Nojiri, G. Polesello, K. Rolbiecki, W.J. Stirling for the 
inspiring collaboration.
The collaboration on this
 work was supported in part by the Polish Ministry of Science and Higher 
Education Grant
1 P03B10830 and by the European Community's Marie-Curie Research 
Training Network 
MRTN-CT-2006-035505 (HEPTOOLS).
}
\vspace*{-1.2cm}

\end{document}